\documentclass{PoS}
\usepackage{amsmath,amssymb}
\usepackage{graphicx}

\def\mc#1{\mathcal{#1}}
\def\eqref#1{{(\ref{#1})}}

\def\L{\Lambda}
\def\S{\Sigma}
\def\s{\sigma}
\def\D{\Delta}
\def\O{\Omega}
\def\GMO{\textrm{GMO}}

\title{Evidence for chiral logarithms in the baryon spectrum}

\ShortTitle{Baryon chiral logs}

\author{\speaker{Andr\'{e} Walker-Loud} \\
        Nuclear Science Division, Lawrence Berkeley National Laboratory, Berkeley, CA 94720 \\
        E-mail: \email{awalker-loud@lbl.gov}}

\abstract{Using precise lattice QCD computations of the baryon spectrum, we present the first direct evidence for the presence of contributions to the baryon masses which are non-analytic in the light quark masses; contributions which are often denoted \textit{chiral logarithms}.  We isolate the poor convergence of $SU(3)$ baryon chiral perturbation theory to the flavor-singlet mass combination.  The flavor-octet baryon mass splittings, which are corrected by chiral logarithms at next to leading order in $SU(3)$ chiral perturbation theory, yield baryon-pion axial coupling constants $D, F, C$ and $H$ consistent with QCD values; the first evidence of chiral logarithms in the baryon spectrum.  The Gell-Mann--Okubo relation, a flavor-\textbf{27} baryon mass splitting, which is dominated by chiral corrections from light quark masses, provides further evidence for the presence of non-analytic light quark mass dependence in the baryon spectrum; we simultaneously find the GMO relation to be inconsistent with the first few terms in a taylor expansion in $m_s - m_l$, which must be valid for small values of this $SU(3)$ breaking parameter.  Additional, more definitive tests of $SU(3)$ chiral perturbation theory will become possible with future, more precise, lattice calculations.
}

\FullConference{ The XXIX International Symposium on Lattice Field Theory - Lattice 2011\\
July 10-16, 2011\\
Squaw Valley, Lake Tahoe, California}

\begin{document}

%
\section{Introduction}
%
Lattice QCD calculations are now performed with light quark masses at or near their physical values~\cite{latt_physPoint}, opening a new era for detailed comparisons with chiral perturbation theory ($\chi$PT).
While this program has been very successful for mesons~\cite{Colangelo:2010et}, the application to baryon properties has been wrought with significant challenges mainly from issues of convergence of the perturbative expansion.
Recent analysis suggests the convergence of the two-flavor expansion for the nucleon mass is limited to $m_\pi \lesssim 300$~MeV~\cite{WalkerLoud:2008bp,WalkerLoud:2008pj}.
The $SU(3)$ chiral expansion has similar but more severe problems.
In heavy baryon $\chi$PT~\cite{hbchipt} (HB$\chi$PT), the small expansion parameter is given by $\epsilon \sim m_K / \Lambda_\chi$, whereas for the pion-octet $\chi$PT, the small expansion parameter is $\epsilon_\phi \sim \epsilon^2$.
Several offshoots of HB$\chi$PT have been developed in an effort to improve the convergence of the theory~\cite{baryon_chipt}.  
We review a new application of an old idea: combining the large $N_c$ expansion with the $SU(3)$ chiral expansion~\cite{largeN}.
This approach has a few formal advantages over the other methods.  In the large $N_c$ limit, there is an extra symmetry, the contracted spin-flavor symmetry allowing for an unambiguous field-theoretic method to include the low lying decuplet baryon resonances in the theory; in the large $N_c$ limit, the spin-$1/2$ and -$3/2$ baryons become degenerate and infinitely heavy.

Having a controlled expansion is necessary but not sufficient to claim success.  The principle prediction from $\chi$PT are the contributions to hadronic observables which are non-analytic in the light quark masses, arising from pion-octet loops, which often contribute $\ln(m_{K,\pi,\eta}^2)$ terms to hadronic observables, and are commonly referred to as \textit{chiral logs}. 
These contributions can not arise from a finite number of local counterterms but only from the long range contributions from the light pion octet degrees of freedom, the \textit{pion cloud}. 
Isolating this predicted light quark mass dependence in lattice QCD results has been a major challenge for many years.  The definitive identification of these contributions is hailed as a signal that the $up$ and $down$ (and $strange$) quarks are sufficiently light that the lattice results can be described accurately by $\chi$PT.
This task has proved to be very challenging, as often, these non-analytic light quark mass contributions are subleading, or masked by other systematics.

We report on the first substantial and direct evidence of the presence of non-analytic light quark mass dependence in the baryon spectrum, work which was performed in Ref.~\cite{WL}.

%
\section{Evidence for non-analytic light quark mass dependence}
%
In Ref.~\cite{Jenkins:1995td}, linear combinations of the ground state baryon spectrum were constructed to isolate various operators in the combined $SU(3)$ and large $N_c$ expansions.  These mass relations were compared with lattice calculations and it was demonstrated the predicted mass hierarchy persisted over a large range of quark masses~\cite{Jenkins:2009wv}.
Here, we focus on three of these mass relations in addition to the Gell-Mann--Okubo relation, and provide evidence for the presence of non-analytic light quark mass dependence in the baryon spectrum.
The heavy baryon Lagrangian was formulated in the $1/N_c$ expansion in Ref.~\cite{Jenkins:1995gc}, providing relations amongst the various LECs.  In particular, the leading quark mass dependent operators satisfy the following relations at subleading order in $1/N_c$
\begin{eqnarray}\label{eq:lgNnlomQ}
b_D &=& \frac{1}{4} b_{2}, \quad
b_F = \frac{1}{2} b_{1} + \frac{1}{6} b_{2}, \quad
b_T = -\frac{3}{2} b_{1} -\frac{5}{4} b_{2}, \quad
\nonumber\\
\s_B &=& \frac{1}{2} b_{1} + \frac{1}{12} b_{2}, \quad
\s_T = \frac{1}{2} b_{1} + \frac{5}{12} b_{2}\, ,
\end{eqnarray}
while the axial couplings satisfy the relations at leading order in $1/N_c$
\begin{equation}\label{eq:lgNAxial}
D = \frac{1}{2}a_{1}, \qquad
F = \frac{1}{3}a_{1}, \qquad
\mc{C} = -a_{1}, \qquad
\mc{H} = -\frac{3}{2}a_{1}\, ,
\end{equation}
significantly reducing the number of LECs to be determined in the analysis.

The numerical data is take from Ref.~\cite{WalkerLoud:2008bp}, which is a mixed-action lattice calculation with domain-wall valence fermions on the dynamical MILC configurations.  While the relevant mixed-action EFT is known~\cite{MAEFT}, the lattice results exist at only a single lattice spacing.
We therefore restrict our analysis to that of the continuum HB$\chi$PT.

%
\subsection{Mass relation $R_1$}
\begin{figure}
\begin{tabular}{cc}
\includegraphics[width=0.48\textwidth]{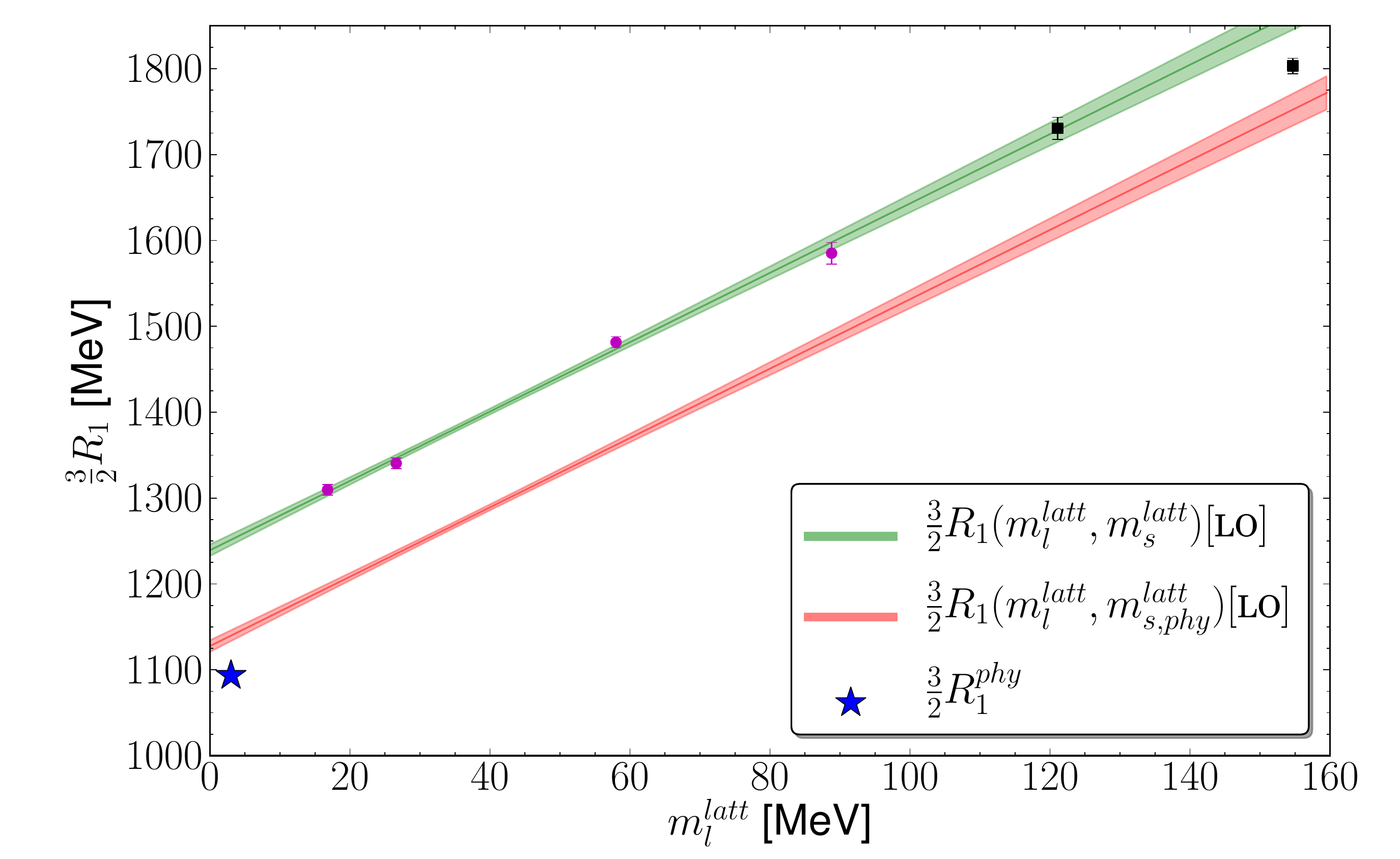}
&\includegraphics[width=0.48\textwidth]{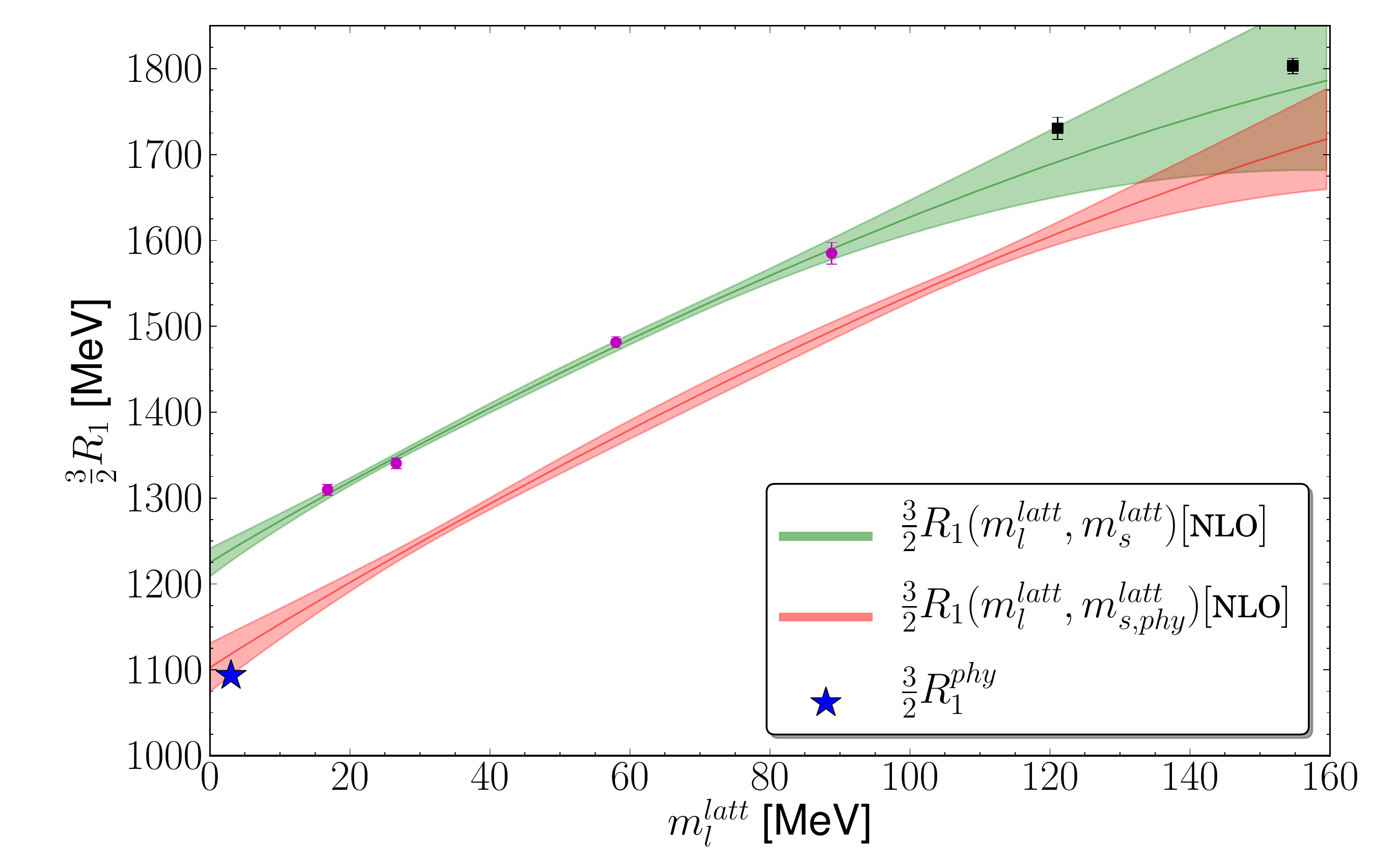}
\end{tabular}
\caption{\label{fig:R1} \textit{Representative fits to $R_1$ from LO (left) and NLO (right) HB$\chi$PT analysis.  The blue star is the physical value, not used in the analysis.}}
\end{figure}

%
We begin with the flavor singlet mass relation $R_1$:
\begin{equation}
R_1 = \frac{25(2M_N + M_\L + 3M_\S + 2M_\Xi) - 4(4M_\D + 3M_{\S^*} + 2M_{\Xi^*} + M_\O)}{240}\, .
\end{equation}
To NLO in the chiral expansion and using the large $N_c$ operator relations~\eqref{eq:lgNnlomQ}, \eqref{eq:lgNAxial},
\begin{eqnarray}
\frac{3}{2}R_1(m_l,m_s) &=& M_0 - \frac{3}{4}\left( b_1 + \frac{5}{18} b_2 \right) (2m_l + m_s)
	-\frac{a_1^2}{96(4\pi f)^2} \bigg[
		35 \left( 3\mc{F}_\pi^0 +4\mc{F}_K^0 + \mc{F}_\eta^0 \right)
\nonumber\\&& \qquad\qquad\qquad
		+50 \left( 3\mc{F}_\pi^\D + 4\mc{F}_K^\D + \mc{F}_\eta^\D \right)
		-4 \left( 3\mc{F}_\pi^{-\D} + 4\mc{F}_K^{-\D} + \mc{F}_\eta^{-\D}\right)
		\bigg]\, ,
\end{eqnarray}
with $\mc{F}_\phi^\D = \mc{F}(m_\phi,\D,\mu)$ defined in Ref.~\cite{WL}, encoding the leading non-analytic light quark mass dependence in the baryon spectrum.  Both LO ($a_1 = 0$) and NLO fits were performed to the lattice data, for a variety of ranges of the light quark masses.  The NLO analysis yielded the LECs
\begin{equation}
M_0[\textrm{NLO}] = 899(40) \textrm{ MeV}, \quad
\left[b_1 + \frac{5}{18}b_2\right][\textrm{NLO}] = -3.26(70), \quad
a_1[\textrm{NLO}] = 0.24(30)\, ,
\end{equation}
Figure~\ref{fig:R1} displays representative fits.  The lower error band is obtained by setting $m_s^{latt} \rightarrow m_{s,phy}^{latt}$, determined from an NLO $\chi$PT~\cite{Gasser:1984gg} analysis of the pion and kaon spectrum.
The small value for the axial coupling, $a_1$ signals a lack of contributions from the non-analytic light quark mass effects, consistent with the $SU(2)$ chiral extrapolation analysis of the nucleon mass~\cite{WalkerLoud:2008bp,WalkerLoud:2008pj}, but inconsistent with their phenomenological determination~\cite{FloresMendieta:1998ii} or direct computation from lattice QCD~\cite{Lin:2007ap}.
One is left to conclude that $SU(3)$ HB$\chi$PT does not provide a controlled perturbative expansion for $R_1$ over the range of quark masses explored in this work.

%
\subsection{Mass relations $R_3$ and $R_4$}
%

\begin{figure}
\begin{tabular}{cc}
\includegraphics[width=0.48\textwidth]{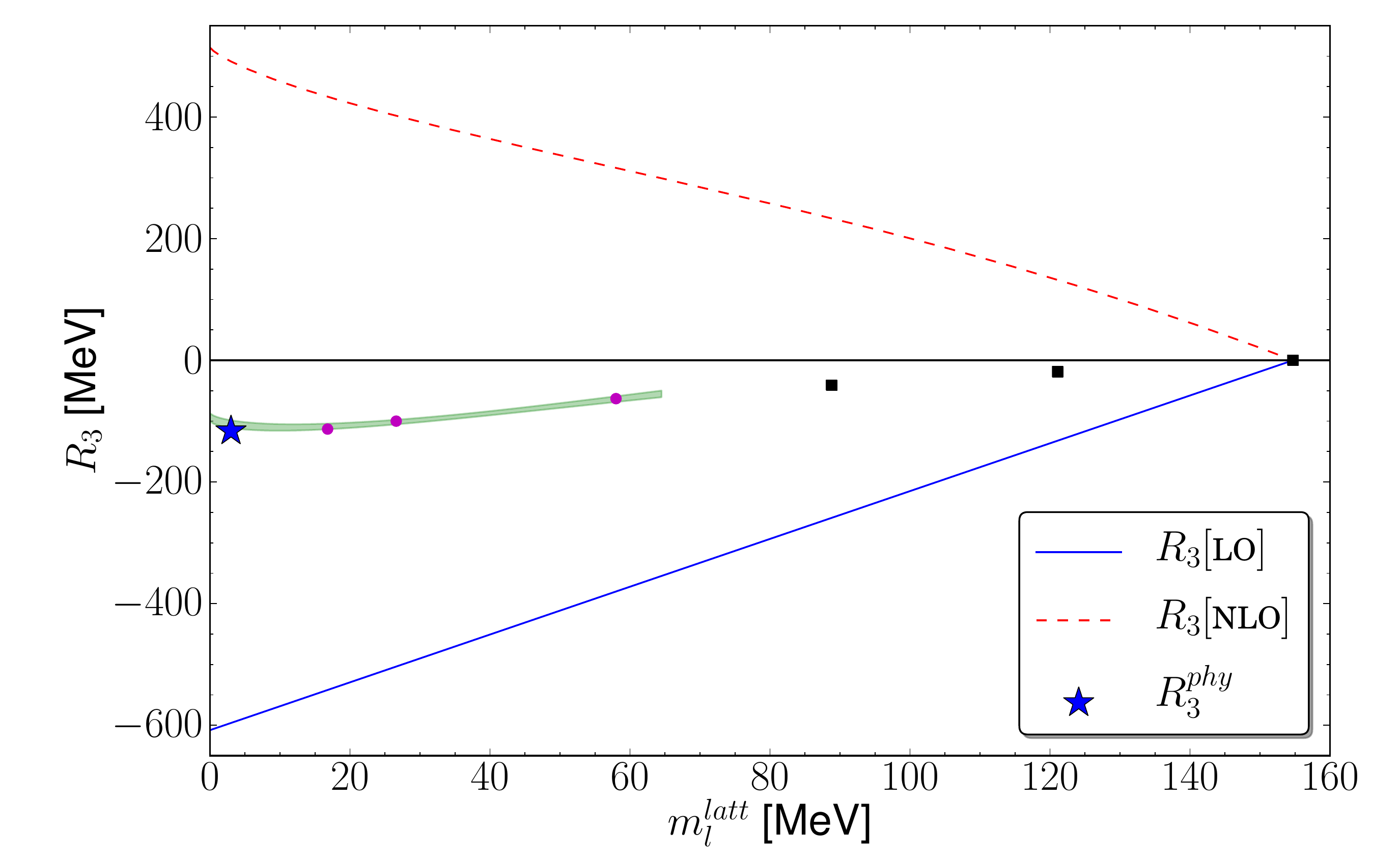}
&\includegraphics[width=0.48\textwidth]{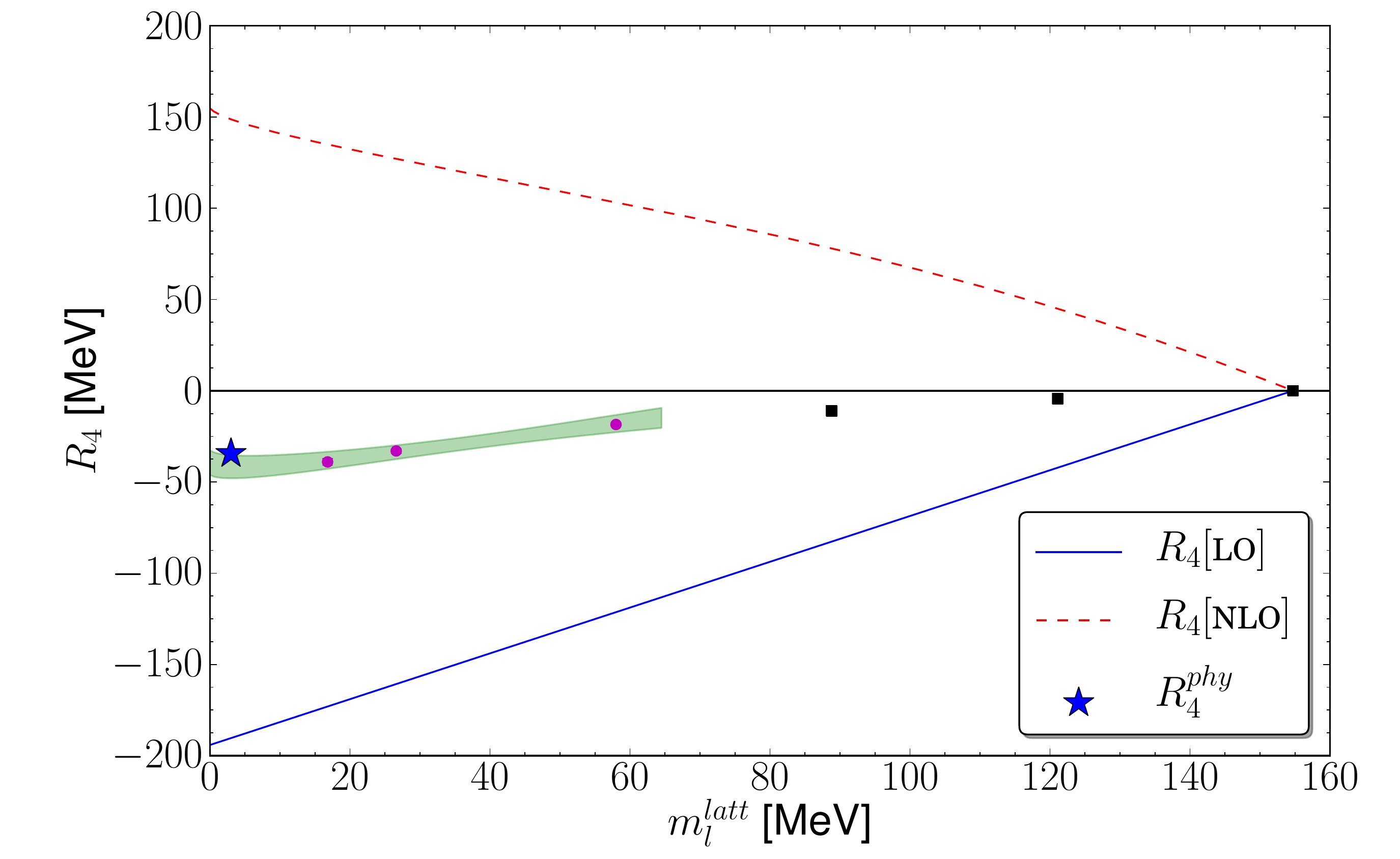}
\end{tabular}
\caption{\label{fig:R3R4_expansion} \textit{The LO and NLO contributions to $R_3$ (left) and $R_4$ (right) from a sample analysis.  A (blue) star is used to denote the physical values, not included in the analysis.}}
\end{figure}

We next examine the flavor-octet mass relations $R_3$ and $R_4$
\begin{eqnarray}
R_3 &=& \frac{5(6M_N + M_\L - 3M_\S -4M_\Xi) - 2(2M_\D -M_{\Xi^*} -M_\O)}{78}
\nonumber\\
R_4 &=& \frac{M_N + M_\L - 3M_\S +M_\Xi}{6}
\end{eqnarray}
These mass relations vanish in both the $SU(3)$ chiral and vector limits, making them more sensitive to the non-analytic light quark mass dependence appearing at NLO in the chiral expansion.
To NLO in the chiral expansion and using the large $N_c$ operator relations~\eqref{eq:lgNnlomQ}, \eqref{eq:lgNAxial},
\begin{eqnarray}
R_3(m_l,m_s) &=& \frac{20}{39} b_1 (m_s - m_l)
	-\frac{a_1^2}{117(4\pi f)^2} \bigg[
		20 \left( 3\mc{F}_\pi^0 - 2\mc{F}_K^0 - \mc{F}_\eta^0 \right)
\nonumber\\ &&\qquad\qquad\qquad
		+35 \left( 3\mc{F}_\pi^\D - 2\mc{F}_K^\D - \mc{F}_\eta^\D \right)
		- \left(3\mc{F}_\pi^{-\D} - 2\mc{F}_K^{-\D} - \mc{F}_\eta^{-\D} \right)
	\bigg]\, ,
\\
R_4(m_l,m_s) &=& -\frac{5}{18}b_2 (m_s - m_l)
	+\frac{a_1^2}{36} \frac{
		\left( 3\mc{F}_\pi^0 - 2\mc{F}_K^0 - \mc{F}_\eta^0 \right)
		-8 \left( 3\mc{F}_\pi^\D - 2\mc{F}_K^\D - \mc{F}_\eta^\D \right)
		}{(4\pi f)^2}\, .
\end{eqnarray}
The LO expressions ($a_1 = 0$) fail to describe the numerical results; it is clear higher order contributions are necessary for the extrapolations of these mass relations.
At NLO, the analysis of $R_3$ and $R_4$ becomes correlated.  The full covariance matrix is constructed as described in Ref.~\cite{Jenkins:2009wv}.  The NLO analysis, considering several possible ranges of $m_l^{latt}$ yields values of the LECs
\begin{equation}
	b_1[\textrm{NLO}] = -6.6(5), \quad
	b_2[\textrm{NLO}] = 4.3(4), \quad
	a_1[\textrm{NLO}] = 1.4(1).
\end{equation}
Using the leading large $N_c$ relations, Eq.~\eqref{eq:lgNAxial}, this corresponds to 
\begin{equation}
	D = 0.70(5), \quad
	F = 0.47(3), \quad
	C = -1.4(1), \quad
	H = -2.1(2).
\end{equation}
The significance of this is prominent; the large value of the axial coupling is strong evidence for the presence of the non-analytic light quark mass dependence in these mass relations.  Further, this is the first time an analysis of the baryon spectrum has returned values of the axial couplings consistent with phenomenology.
\footnote{It is interesting to note that while the $SU(3)$ chiral expansion for the baryon spectrum is not convergent, it was found that the volume dependence of the octet baryon masses is consistent with $SU(3)$ HB$\chi$PT.  Analysis of the volume dependence yielded a large value of $g_{\pi N\Delta}\ (C)$ with $g_A$ fixed to its physical value~\cite{Beane:2011pc}.} 
However, caution is in order.  Examining the resulting contributions to $R_3$ and $R_4$ from LO and NLO separately, one observes a delicate cancellation between the different contributions, see Fig.~\ref{fig:R3R4_expansion}.  Further studies are needed with more numerical data sufficient to also constrain the sub-leading large $N_c$ axial coefficient $a_2$ as well as the NNLO contributions.

%
\subsection{Gell-Mann--Okubo Relation}
%

The last mass relation we study is the flavor-\textbf{27} Gell-Mann--Okubo relation
\begin{equation}\label{eq:GMO}
\D_{\GMO} = \frac{3}{4}M_\L + \frac{1}{4}M_\S - \frac{1}{2}M_N - \frac{1}{2}M_\Xi\, .
\end{equation}
The first non-vanishing contribution to $\D_\GMO$ comes from the NLO chiral loops, which are non-analytic in the light quark masses.  For this reason, the GMO relation is of particular interest to study with lattice QCD.  We extend the previous analysis~\cite{Beane:2006pt,WalkerLoud:2008bp} in a few important ways.  
Close to the $SU(3)$ vector limit, the GMO relation can be described by a taylor expansion in $m_s - m_l$,
\begin{equation}\label{eq:GMOnnloV}
	\D_{\GMO}^V = d_2 \left(m_s - m_l \right)^2 
		+ d_3 \left(m_s - m_l \right)^3
		+ \cdots
\end{equation}
The leading term proportional to $(m_s - m_l)$ must vanish as it transforms as a flavor-$\mathbf{8}$.  The first non-vanishing contribution is equivalent to a next-to-next-to-leading order (NNLO) contribution from HB$\chi$PT and the $(m_s - m_l)^3$ contribution is equivalent to an NNNNLO HB$\chi$PT contribution.
We demonstrate these first few terms in the Taylor expansion about the $SU(3)$-vector limit are inconsistent with the lattice data as $m_l^{latt} \rightarrow 0$.
We extend the previous analysis to include the NNLO HB$\chi$PT contributions, with the axial couplings constrained by the analysis of $R_3$ and $R_4$.  It is found only NNLO HB$\chi$PT, which is dominated by the non-analytic light quark mass contributions, can naturally accommodate the strong light quark mass dependence observed in the numerical results.
At NLO in the chiral expansion and using the large $N_c$ operator relations~\eqref{eq:lgNnlomQ}, 
\begin{equation}
\D_\GMO[\textrm{NLO}] = \frac{a_1^2}{36(4\pi f)^2}
	\left[ \mc{F}_\pi^0 - 4\mc{F}_K^0 + 3\mc{F}_\eta^0
		+2\left( \mc{F}_\pi^\D - 4\mc{F}_K^\D + 3\mc{F}_\eta^\D \right)
	\right]\, .
\end{equation}
The full NNLO formula, determined from Ref.~\cite{WalkerLoud:2004hf} can be found in Ref.~\cite{WL}.

\begin{figure}
\includegraphics[width=0.98\textwidth]{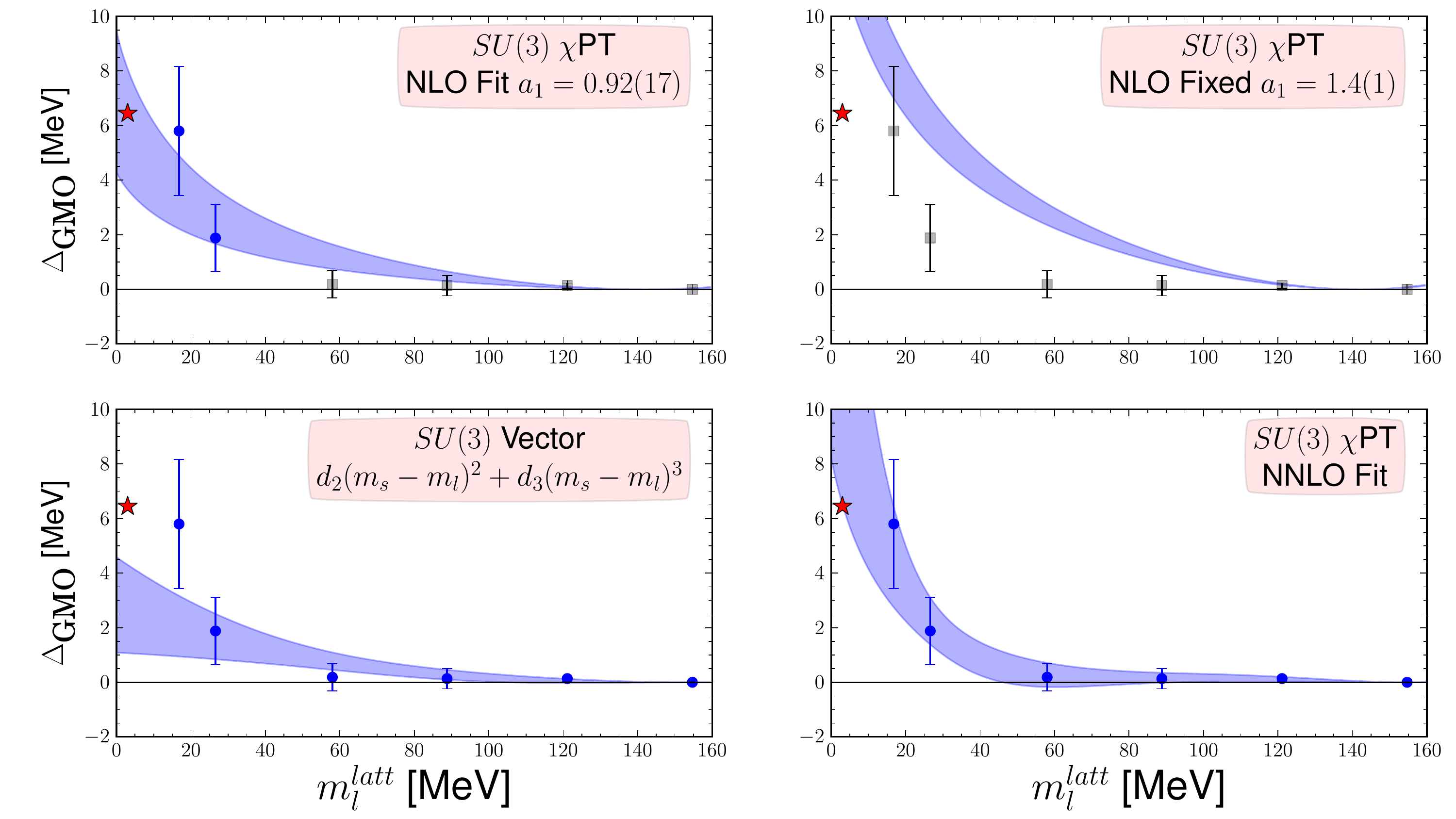}
\caption{\label{fig:GMO} \textit{GMO mass splitting plotted as a function of $m_l^{\rm latt}$.  The $*$ is the PDG point, not included in the analysis.
The various fits are described in the text.
In a given plot, the filled (blue) circles denote results included in the analysis while the open (gray) boxes are excluded.}}
\end{figure}
In Fig.~\ref{fig:GMO}, four plots are displayed.  The first plot (upper left) is the result of an NLO analysis of the GMO formula, allowing the axial coupling to be determined from the data, resulting in a small, but non-zero value for $a_1$.
The second plot (upper right) displays the predicted value of the GMO relation from NLO taking the determination of $a_1$ from the analysis of $R_3$ and $R_4$.
The third plot (bottom left) shows the result of a taylor expansion about the $SU(3)$ vector limit fitting the first two non-vanishing terms.
Finally, the NNLO analysis is displayed, using the value of $a_1$ determined from $R_3$ and $R_4$ (bottom right).
Only the NNLO analysis is consistent with the values of the numerical data over the full range of light quark masses, in particular, the steep rise observed as $m_l^{latt} \rightarrow 0$, as well as the value of the axial coupling $a_1$ determined from phenomenology.
This is further evidence for non-analytic light quark mass dependence in the baryon spectrum.

%
\section{Conclusions}
%
We have presented the first substantial evidence for the presence of non-analytic light quark mass dependence in the baryon spectrum, with further analysis details in Ref.~\cite{WL}.
This was achieved by comparing the predictions from HB$\chi$PT combined with the large $N_c$ expansion to relatively high statistics lattice computations of the octet and decuplet baryon spectrum.
An analysis of mass relations $R_3$ and $R_4$ provided for the first time, values of the axial couplings which are consistent with the phenomenological determination, signaling significant contributions from non-analytic light quark mass dependence in $R_3$ and $R_4$: utilizing the leading large $N_c$ expansion,
\begin{center}
$D = 0.70(5)\, ,\quad
F = 0.47(3)\, ,\quad
C = -1.4(1)\, ,\quad
H = -2.1(2)\, .$
\end{center}
It was further demonstrated that the Gell-Mann--Okubo relation is inconsistent with the first two non-vanishing terms in a taylor expansion about the $SU(3)$ vector limit, and that the steep rise in the numerical data, observed as $m_l^{latt} \rightarrow 0$, can only be described by the NNLO heavy baryon $\chi$PT formula which is dominated by chiral loop contributions.
Taken together, these observations indicate the first significant evidence for the presence of non-analytic light quark mass dependence in the baryon spectrum.

However, there are several known systematics which were not addressed in the present article, and require future, more precise lattice results:
the numerical data used~\cite{WalkerLoud:2008bp} exist at only a single lattice spacing:
continuum $\chi$PT analysis was performed:
there may be contamination from finite volume effects~\cite{Beane:2011pc}:
the convergence issues need further examination:
more precise numerical results are needed to explore mass relations $R_5$ -- $R_8$ which are more sensitive to non-analytic light quark mass dependence:
results with smaller values of the light quark mass are desirable:
the strange quark mass used in this work is known to be $\sim$25\% to large~\cite{Aubin:2004ck}.

\end{document}